\documentclass[%
preprint,
 amsmath,amssymb,
 aps,
prl,
endfloats*,
]{revtex4-1}

\usepackage{rotating}
\usepackage{lipsum}
\usepackage{graphicx}% Include figure files
\usepackage{dcolumn}% Align table columns on decimal point
\usepackage{bm}% bold math
\usepackage{changes}
\definechangesauthor[name={HI}, color=orange]{HI}
\begin{document}
\setlength {\marginparwidth} {2cm}
%\preprint{APS/123-QED}

\title{Combining quantum spin hall effect and superconductivity in few-layer stanene}% Force line breaks with \\
%\thanks{A footnote to the article title}%

\author{Chenxiao Zhao$^{1}$, Jin Qin$^{1}$, Bing Xia$^{1}$, Bo Yang$^{1}$, Hao Zheng$^{1,2,3}$, Shiyong Wang$^{1,2,3}$,  Canhua liu$^{1,2,3}$, Yaoyi Li$^{1,2,3}$, Dandan Guan$^{1}$,
Jinfeng Jia$^{1,2,3}$\footnote{jfjia@sjtu.edu.cn}}
\affiliation{$^{1}$Key Laboratory of Artificial Structures and Quantum Control (Ministry of Education), Shenyang National Laboratory for Materials Science, School of Physics and Astronomy, Shanghai Jiao Tong University, Shanghai 200240, China}
\affiliation{$^{2}$Tsung-Dao Lee Institute, Shanghai Jiao Tong University, Shanghai 200240, China}
\affiliation{$^{3}$CAS Center for Excellence in Topological Quantum Computation, University of Chinese Academy of Sciences, Beijing 100190, China}

\date{\today}% It is always \today, today,
             %  but any date may be explicitly specified

\begin{abstract}
\textbf{
Stanene was proposed to be a quantum spin hall insulator containing topological edges states and a time reversal invariant topological superconductor hosting helical Majorana edge mode. Recently, experimental evidences of existence of topological edge states have been found in monolayer stanene films and superconductivity has been observed in few-layer stanene films excluding single layer. An integrated system with both topological edge states and superconductivity are higly pursued as a possible platform to realize topological superconductivity. Few-layer stanene show great potential to meet this requirement and is highly desired in experiment. Here we successfully grow few-layer stanene on bismuth (111) substrate. Both topological edge states and superconducting gaps are observed by in-situ scanning tunneling microscopy/spectroscopy (STM/STS). Our results take a further step towards topological superconductivity by stanene films.   
}
\end{abstract}

%\pacs{Valid PACS appear here}% PACS, the Physics and Astronomy
                             % Classification Scheme.
%\keywords{Suggested keywords}%Use showkeys class option if keyword
                              %display desired
\maketitle

%\tableofcontents

\textbf{Introduction}

Stanene, or called monolayer $\alpha$-Sn (111), is the tin analogue of graphene. It has a distorted honeycomb structure composed of two sub-lattices of Sn atoms. Stanene is proposed to be a large gap 2D topological insulator (TI)\cite{Xu2013}, which is characterized by dissipationless conducting helical edge channels protected by bulk band topology \cite{Hasan2010,Qi2011}. However, the band structure and topological properties of stanene are tremendously rely on lattice constants, chemical functionality and layers thickness\cite{Xu2015,Gou2017,wang2015,Chou_2014,yun2017,liu2019}. This sensitivity of topological properties, although makes stanene to be a tunable topological material, increases the difficulties for experimental verifications. Stanene films grown on different substrates exhibite distinct properties. In some experiments, stanene films show metallic band structures such as stanene / Bi$_2$Te$_3$ (111) \cite{zhu2015}, or stanene /Sb (111) \cite{Gou2017}. In some other experiments, it shows a trivial band gap, such as stanene grown on PbTe (111) and InSb (111) \cite{caizhi2018,zang2018}. A breakthrough in nontrivial band topology is achieved when growing stanene on Cu (111) surface, where an inverted band is observed\cite{deng2018}. But stanene films on Cu surface are stable only at low temperature, and structural distortions spontaneously happen at room temperature\cite{deng2018}. Recently, stanene films grow on InSb (111) also show the existence of edge states\cite{Zheng_2019}, but a vast amount of defects are formed on its surface which is unfavorable for multilayer growth. Up to now, few-layer stanene with topological edge states have not been successfully grown. 

Stanene is also predicted to be a topological superconductor (TSC) once breaking inversion symmetry\cite{qi2009}. TSC is characterized by a fully superconducting gap with triplet pairing in the bulk and self-conjugate excitations at their boundaries\cite{sato2017,Qi2011}. This kind of boundary excitation is called Majorana mode, which has huge potential in topological quantum computation (TQC)\cite{beenakker2013,sau2010,alicea2011}. Previously, TSCs are commonly realized by using proximity effect by a combined system involving s-wave superconductors (SCs), such as 3D TI/SC heterostructure\cite{fu2008,jinpeng2015}, iron chains on SC\cite{nadjperge2014} and semiconductor chains with large spin orbital coupling on SC\cite{mourik2012}. These combined systems are complicated to fabricate. Recently, iron based superconductor, like FeTe$_{0.55}$Se$_{0.45}$ \cite{wang2018}and (Li $_{0.84}$Fe$_{0.16}$ )OHFeSe\cite{Liu2018}, are proposed to be intrinsic TSC. But these materials have multi-components and require precise control of the ratio between different elements, which increases difficulties for sample growth. Stanene, as a single element material, is more feasible to acquire than the other candidates of TSC. At their edges, the helical Majorana modes exists which probably can also be used for TQC\cite{liuxiongjun2014,liuxiongjun2016}. What is more, the TSC based on stanene is time reversal invariant, and the presence of Majorana modes do not need the participation of magnetic field or magnetic atoms. These all make stanene to be a stable and feasible platform for TQC. Recently, the superconductivity of few-layer stanene have been measured \cite{liao2018superconductivity}and a newly proposed pairing mechanism, called type II Ising pairing\cite{Wangchong2019}, are prosed to interpret its anomalous large in-plane upper critical field\cite{falson2020}. The topological properties of its superconductivity is still not clear and the experimental verifications are also lacking.

Here we grow few-layers stanene on Bi (111) substrate. The topological edge states are clearly observed at film edges of 2-4 layer stanene. And the superconductivity is detected in both monolayer and multilayer stanene films. This results makes few-layer stanene a good platform to realize topological superconductor.

%%%%%%%%%%%%%%%%%%%%%
%%%%%%%%%%%%%%%%%%%%%
%END INTRODUCTION, BEGIN EXP.
%%%%%%%%%%%%%%%%%%%%%
%%%%%%%%%%%%%%%%%%%%%

\textbf{On the quantum spin hall effet}

%%%%%%%%%%%%%%%%%%%%%
%%%%%%%%%%%%%%%%%%%%%
%FIGURE 1: 
%%%%%%%%%%%%%%%%%%%%%
%%%%%%%%%%%%%%%%%%%%%

\begin{figure}
\includegraphics[width=15cm]{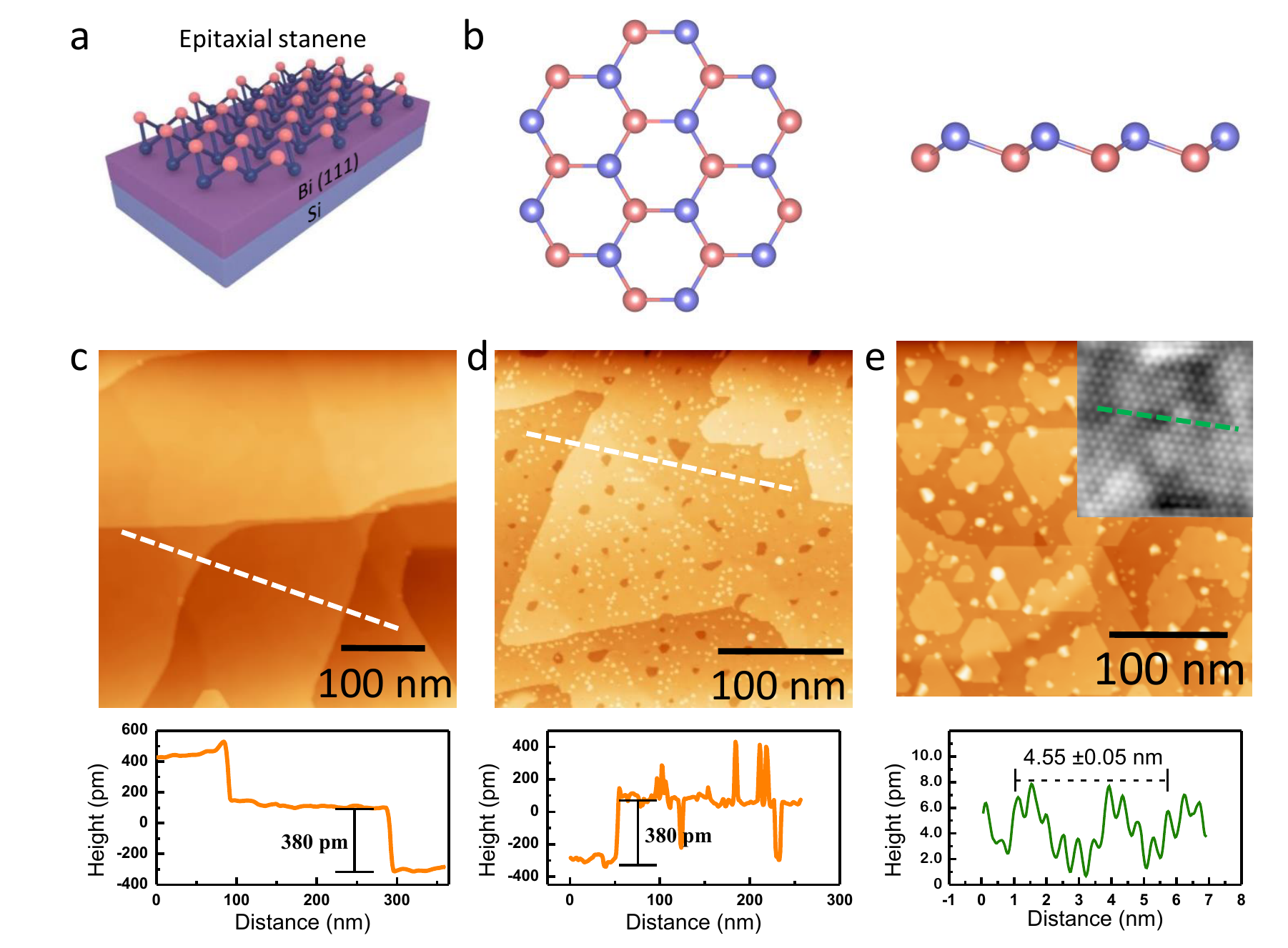}
\caption{\label{fig1}
Topography of Bi (111) substrate and first-layer stanene. a, Schematic diagram of sample structure: stanene/Bi(111)/Si. b, Top and side views of monolayer stanene. Two sublattices are represented by different colors. c, Topography of Bi (111) substrate. The profile of white dotted line is shown at the bottom. d, Topography of first layer stanene. The profile of white dotted line is shown at the bottom. e, Topography of 3.5 layer stanene films. The inset shows atomic resolved image of stanene film and the lattice distances are shown at the bottom.
}
\end{figure}

The Bi (111) surface is chosen to be the substrate to grow stanene films because both of them have a hexagonal lattice and its lattice constant ($\sim$0.454 nm) is close to that of stanene ($\sim$0.467 nm). The sample structure is shown in Fig.1a. The bismuth layer is as thick as ~10 nm to eliminate the epitaxial strain with silicon wafer and to acquire large flat terraces, which are in favor of growing stanene flims. Fig.1b shows the structure of monolayer stanene containing two sub-layers of Sn atoms. Two sub-lattices form a buckled honeycomb structure. The buckled configuration insures the stability of monolayer stanene. The large Bi (111) films are shown in Fig.1c. A cut line crossing two steps of Bi (111) is shown in the lower panel, where the step height is ~380 pm. The topographic image of monolayer stanene is shown in Fig. 1d, with coverage more than 90 $\%$. There is no buffer layer between stanene and Bi(111) and the stanene films are large and flat without distortions as grown on Bi$_2$Te$_3$ (111)\cite{zhu2015}. Some Sn clusters accumulate at the surface of stanene films, but it can be dramatically reduced by annealing process (see supplementary material for details). Also, a line profile is shown at the bottom, where the height of stanene is ~ 380 pm, similar with that of Bi substrate. It seems that Bi (111) is very suitable to growth stanene and we can grow multilayer stanene on it. For 3.5 layer stanene films, the step height of stanene is also $\sim$380 pm, and some Sn clusters grow larger and perfer to reside at the film edges (see the lighter spots in Fig.1e). The atomic resolved image of stanene is shown in the inset and the upper sublattice Sn atoms construct a hexagol lattcie with a constant of $\sim$0.455 nm, which is the same with Bi (111). High resolution image show that the stacking manner of multilayer stanene indeed follows the structure of $\alpha$-Sn (see supplemenatry materials).

%%%%%%%%%%%%%%%%%%%%%
%%%%%%%%%%%%%%%%%%%%%
%FIGURE 2: 
%%%%%%%%%%%%%%%%%%%%%
%%%%%%%%%%%%%%%%%%%%%

\begin{figure}
\includegraphics[width=15cm]{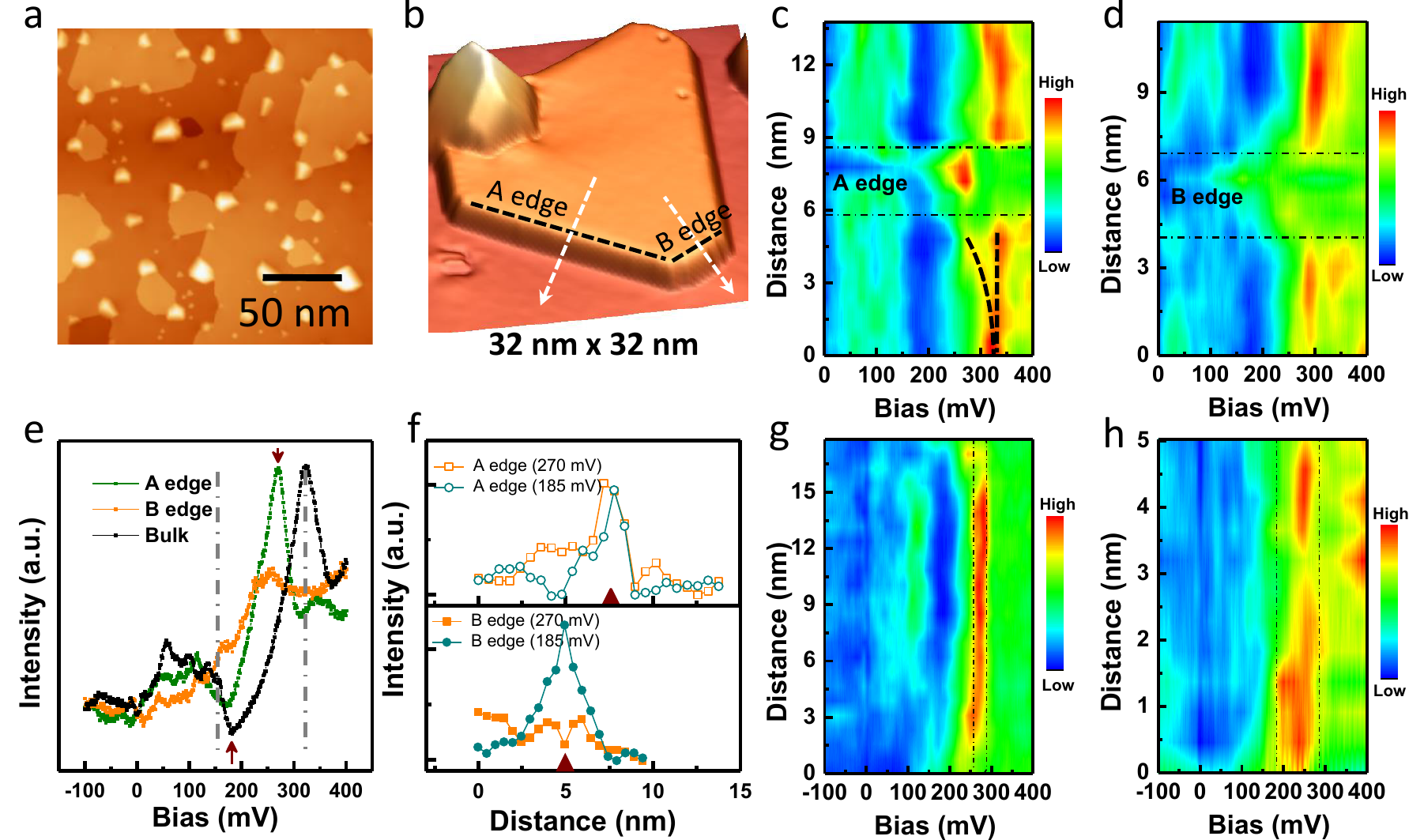}
\caption{\label{fig2} 
dI/dV spectra evidences of edge states. a, Large scale image of 3.5 layer films of stanene. b, A small island of 4th-layer stanene. c, Color image of a series of point spectra along white arrow crossing A edge in b. d, Color image of a series of point spectra along white arrow crossing B edge in b. e, Point spectra at bulk and two different edges of stanene island. f, Spatial variation of intensity of edge states extracted from c (upper) and d (lower). The positions of A and B edges are marked out by red triangular symbol. g and h, Color image of a series of point spectra along dotted lines along A and B edge in b.
}
\end{figure}

The hallmark of QSH system are the topological protected helical conducting edge channels, which are manifested by the enhanced intensity of density of states (DOS) at edges in STM studies. We systematically studied the edge states of different layer stanene films. For the first layer stanene contacting with Bi (111), no uniform edge state is observed. While on 2-4 layer stanene films, significant evidences of edge states have been found. Here, the 4th layer stanene islands are chosen to characterize the edges states (see supplementary material for edges states of different layers). Firstly, the large-scale topographic image of 3.5 layer stanene is shown in Fig.2a, where many islands of 4th layer stanene can be observed. Besides the islands, Sn clusters also exist, which always reside at the edges of stanene island. A hexagonal stanene island is shown in Fig.2b of which the height is also ~380 pm. Two kinds of edges are observed: the longer edges corresponding to the Zigzag edge consisting of upper Sn atoms (marked by A) and the shorter edges corresponding to the Zigzag edge consisting of lower Sn atoms (marked by B). The typical differential conductance (dI/dV) spectra taken at bulk and two edges are shown in figure 2e. The bulk-spectra shows a dip feature ranging from 160mV to 323mV (in between gray dashed lines in Fig. 2e) and the intensity minimum locates at ~185 mV. The absence of insulating gap indacates that the few-layer stanene on Bi (111) is a metallic system. The spectra taken at two edges are distinct. At A edge, a peak emerges at ~ 270 mV, which is in the dip range of the bulk spectrum. It can be attributed to either the emergency of edge states or the energy shift of bulk states. To further investigate the electronic states at A edge, a series of point spectra crossing A edge are taken (along the longer white arrow in Fig.2b) and the corresponding color image is shown in Fig.2c. According to color image, although the bulk peak emerging at ~323 mV show a splitting trend, both of the splitting peaks vanishes near the edges. And a new peak emerges at the edge abruptly. Thus the new emerging peak is attributed to edges states and not the evolutionary shift of bulk states.  

Different from A edge, there is no obvious peak in spectrum taken at B edge, but an overall enhancement of DOS inside the bulk dip range (see the orange spectrum in Fig.2e). The color image of  spectra crossing B edge are shown in Fig.2d. The enhanced DOS at B edges can be observed between two horizontal black dashed lines, where the bulk peaks are absent. The bulk peak also show a splitting on the island, but the splitted peak has a higer energy, in contrast to the splitting along A edge. 

The spatial variations of edge states are extracted from the color images to investigate its spatial distributions. The energies selected are 185 mV and 270 mV (see Fig.2f). At 185mV, both edges show the dramatic increase of DOS, indicating the existence of edge states. While at 270 mV, only A edge shows enhancement of DOS. The above results show that both edges have edge states but the energy distributions are different. The homogeneity of edges states are also investigates. Fig.2g and 2h are color images of spectra taken along A and B edges, respectively (along the black dotted lines in Fig.2b). The peak of A edge states at ~270 meV present regularly along the whole edge until the corner. The spectra color image along B edge also exhibit features of edge states, which is less regular than A edge, probably due to the limited length of B edges. 
\\

%%%%%%%%%%%%%%%%%%%%%
%%%%%%%%%%%%%%%%%%%%%
%FIGURE 3: 
%%%%%%%%%%%%%%%%%%%%%
%%%%%%%%%%%%%%%%%%%%%

\begin{figure}
\includegraphics[width=15cm]{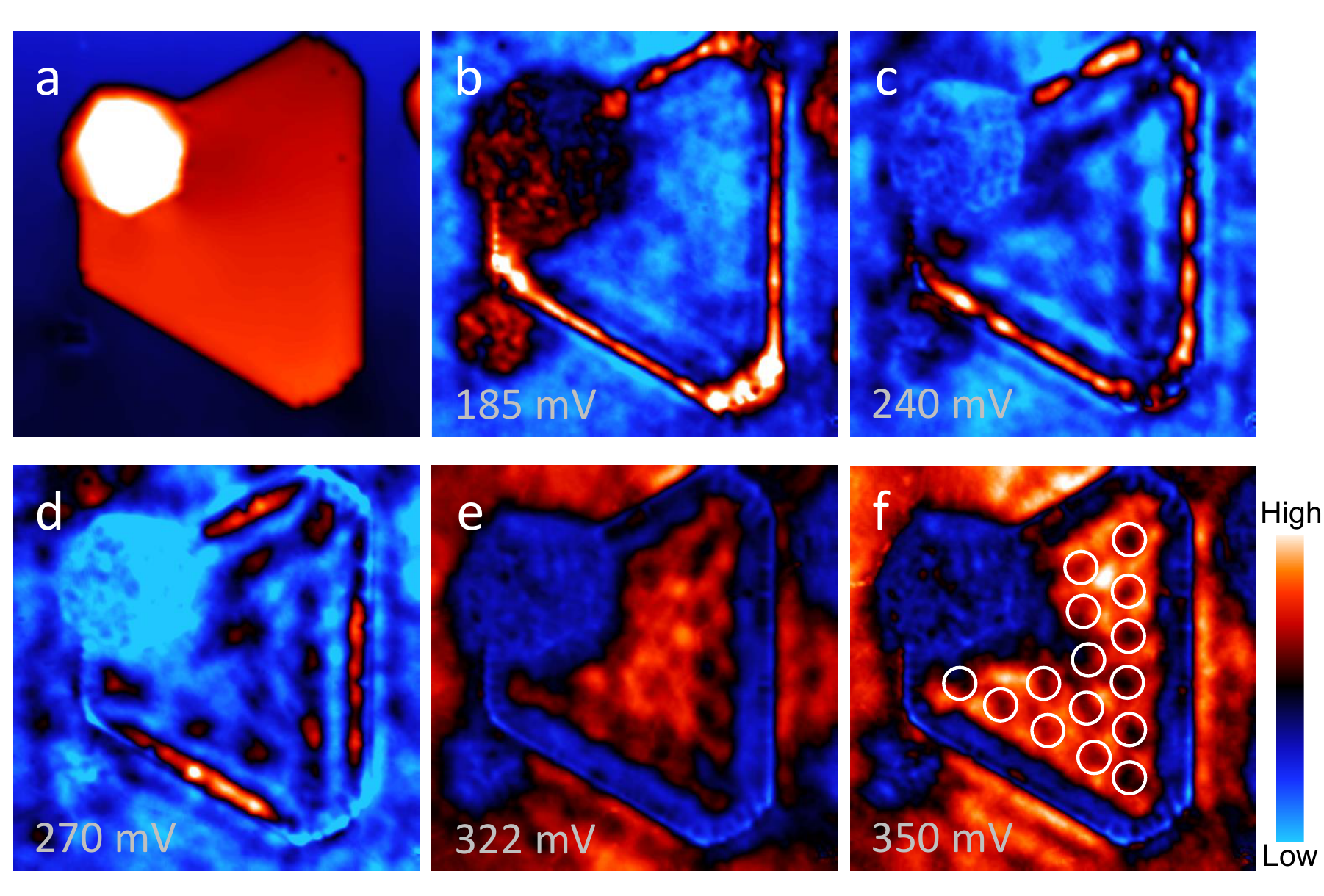}
\caption{\label{fig3}
dI/dV maps of stanene islands. a, Topography of small stanene island  (32 nm $\times$ 32 nm). b-f,dI/dV maps of the same island with different bias voltages.
}
\end{figure}

The dI/dV maps are also used to further investigate the energy and spatial distribution of edge states. The topographic image of the island is shown in Fig. 3a as a reference. A series of dI/dV maps are taken at bias ranging from 185mV to 350 mV. At the bias of 185 mV, the DOS minimum in the bulk spectra, the edge of stanene island is high-lighted due to emergence of edge states. Increasing bias voltage to 240 mV, the highlighted edge contour still exist, but with some breaking points. And some faint interferences pattern emerge in the bulk area. At the bias of A edge peaks ($\sim$270 mV), only A edges are highlighted, the intensity of electronic states at B edges is as weak as the bulk area, consisting with the tunneling spectra. Besides, clear interference patterns (light points inside the island) can be observed. This interferences should be originated from the surface states, thus both A edge states and surface sates are involved in this energy range. Increasing bias to the bulk peak ($\sim$323 mV), the highlighted areas are inside the island and at the lower layer films. The constructive interferences pattern now change into destructive interference pattern which are manifested by dark holes. At last, the bias is increased to 350 mV. The map at this bias are much similar with the the map at 322 mV, but the interference pattern of surface states is more clear (marked by circles).  \\

Based on above experimental results, few-layer stanene is probably a topological insulator with a sememetallic band structure, similar with Bi(111). Direct band gaps may exist at every k point, but there must be overlaps between conduction and valence band in the whole k space, leading the system a semimetal. Thus the edge states overlap with bulk states and surface states. An theoretical verification is still needed to further confirm its topological properties.
%%%%%%%%%%%%%%%%%%%%%
%%%%%%%%%%%%%%%%%%%%%
%FIGURE 4: 
%%%%%%%%%%%%%%%%%%%%%
%%%%%%%%%%%%%%%%%%%%%

\textbf{On the superconductivity}

\begin{figure}
\includegraphics[width=15cm]{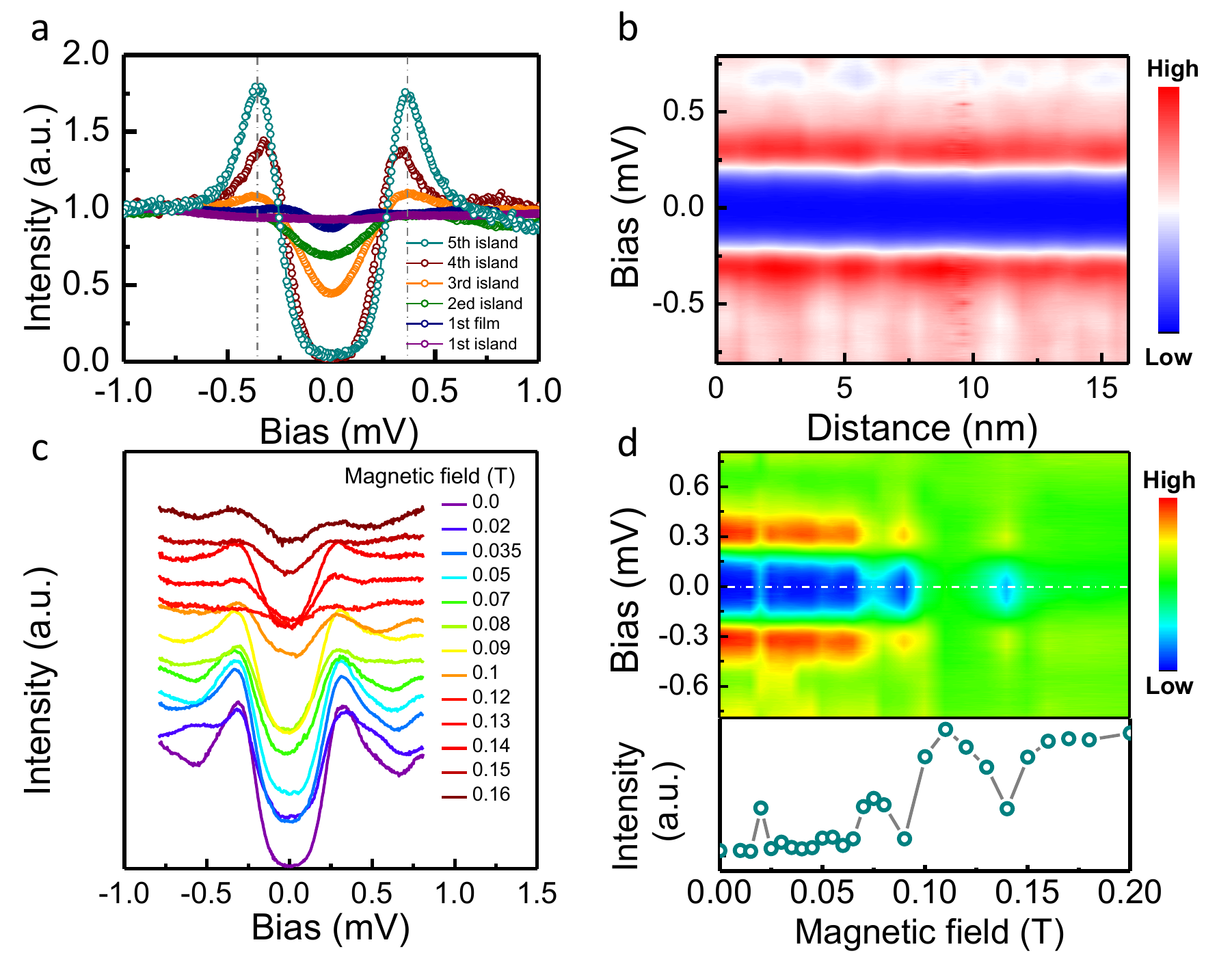}
\caption{\label{fig4}
Superconductivity of stanene films. a, Layer-dependent superconducting spectra of stanene films. b, Color map of a series of dI/dV maps along the same line with figure 2c. c, Superconducting spectra under different magnetic field, all spectra are taken at same position (center of the island shown in figure 2b). There are a total of 27 spectra, only a part of them are shown to give a clear view. The whole spectra are shown in supplementary material. d, Upper panel: The color map of dI/dV spectra with different magnetic field. Lower panel: the DOS intensities at zero bias are extracted to show the fluctuation of superconducting gap with magnetic field.
}
\end{figure}
Another necessary factor to realize time reversal invariant TSC is superconductivity. Although bulk $\alpha$-Sn is not a superconductor\cite{liao2018superconductivity},thin films of $\alpha$Sn exhibit remarkable superconducting signal in transport experiment\cite{liao2018superconductivity,falson2020}. Here, the superconductivity of stanene is confirmed by dI/dV spectra, and the superconducting gaps show significant layer-dependence (shown in Fig.4a). It should be noted that the first layer stanene do not show superconductivity unless the separated small islands develop into a whole film (see supplementary for details). This distinct resusult from ref.\cite{liao2018superconductivity} is probabaly due to different electron doping level by Bi(111) substrate. With increasing the thickness of stanene films, the superconducting gap gets larger and deeper. Fully gaps against thermal excitations at 400 mK (effective temperature is $\sim$ 800 mK) are obtained until the 4th layer. The superconducting gap of stanene film can not be well fited by BCS theory for the present of very sharp cohenrence peaks (see supplmentary materials). Figure 4b shows the spatial homogeneity of superconductivity when crossing the edge of a stanene island (the same one with Fig.2). There is no differences of superconducting gaps between the island and the lower layer film. The uniform superconducting intensity on and under the island may caused by the large coherence length of Sn. However, it seems no helical Majorana mode emerge at the 4th island edges.   

When applying out of plane magnetic field, the superconducting gap shows an anomalous fluctuation.The superconducting gaps do not descend monotonically but with an fluctuation form (see figure 4c and 4d). The density of states at zero bias under different magnetic fields are extracted and shown in figure 4d, where the fluctuation is clearly visible. Considering the magnitude of applied field, its may caused by Little-parks effect\cite{Vaitiekenaseaav3392} which has been recently observed in another topological superconducting system. But a more systematic studies is still needed to explain this phenomenon.

%%%%%%%%%%%%%%%%%%%%%
%%%%%%%%%%%%%%%%%%%%%
%CONCLUSION
%%%%%%%%%%%%%%%%%%%%%
%%%%%%%%%%%%%%%%%%%%%

\textbf{Summary}

The edge states of few-layer stanene are clearly observed by STM/STS. And its spatial distribution is also studied. Besides, the layer-dependence of superconductivity of stanene is characterized. And an anomalous fluctuation of superconducting intensity with magnetic field is observed, which is probably caused by the Little-parks effect. Our results indicate that few-layer stanene is probabaly a good candidate for time reversal invariant TSC.

%\appendix

%\textbf{Methods}

% The \nocite command causes all entries in a bibliography to be printed out
% whether or not they are actually referenced in the text. This is appropriate
% for the sample file to show the different styles of references, but authors
% most likely will not want to use it.
%\nocite{*}

\bibliography{Stanene}% Produces the bibliography via BibTeX.

\end{document}